\documentclass[aps,prb,amsmath,amssymb,showpacs,superscriptaddress,12pt]{revtex4-1}

\usepackage{graphicx}
\usepackage{amsmath,amssymb,amsfonts}
\usepackage{textcomp}
\usepackage{hyperref}
\usepackage{gensymb}
\usepackage{verbatim}
\usepackage{amsmath}
\hypersetup{breaklinks=true,colorlinks=true,urlcolor=black}
\usepackage{color}
\usepackage{soul}
\DeclareGraphicsExtensions{.jpg,.pdf,.png,.eps}

\begin{document}
\title{Manipulating of magnetism in the topological semimetal EuCd$_{2}$As$_{2}$}

\author{Na Hyun Jo}
\thanks{These two authors contributed equally}
\affiliation{Ames Laboratory, Iowa State University, Ames, Iowa 50011, USA}
\affiliation{Department of Physics and Astronomy, Iowa State University, Ames, Iowa 50011, USA}

\author{Brinda Kuthanazhi}
\thanks{These two authors contributed equally}
\affiliation{Ames Laboratory, Iowa State University, Ames, Iowa 50011, USA}
\affiliation{Department of Physics and Astronomy, Iowa State University, Ames, Iowa 50011, USA}

\author{Yun Wu}
\affiliation{Ames Laboratory, Iowa State University, Ames, Iowa 50011, USA}
\affiliation{Department of Physics and Astronomy, Iowa State University, Ames, Iowa 50011, USA}

\author{Erik Timmons}
\affiliation{Ames Laboratory, Iowa State University, Ames, Iowa 50011, USA}
\affiliation{Department of Physics and Astronomy, Iowa State University, Ames, Iowa 50011, USA}

\author{Tae-Hoon Kim}
\affiliation{Ames Laboratory, Iowa State University, Ames, Iowa 50011, USA}

\author{Lin Zhou}
\affiliation{Ames Laboratory, Iowa State University, Ames, Iowa 50011, USA}

\author{Lin-Lin Wang}
\affiliation{Ames Laboratory, Iowa State University, Ames, Iowa 50011, USA}

\author{Benjamin G. Ueland}
\affiliation{Ames Laboratory, Iowa State University, Ames, Iowa 50011, USA}
\affiliation{Department of Physics and Astronomy, Iowa State University, Ames, Iowa 50011, USA}

\author{Andriy Palasyuk}
\affiliation{Ames Laboratory, Iowa State University, Ames, Iowa 50011, USA}

\author{Dominic H. Ryan}
\affiliation{Department of Physics, McGill University, Montreal, Québec H3A 2T8, Canada}

\author{Robert J. McQueeney}
\affiliation{Ames Laboratory, Iowa State University, Ames, Iowa 50011, USA}
\affiliation{Department of Physics and Astronomy, Iowa State University, Ames, Iowa 50011, USA}

\author{Kyungchan Lee}
\affiliation{Ames Laboratory, Iowa State University, Ames, Iowa 50011, USA}
\affiliation{Department of Physics and Astronomy, Iowa State University, Ames, Iowa 50011, USA}

\author{Benjamin Schrunk}
\affiliation{Ames Laboratory, Iowa State University, Ames, Iowa 50011, USA}
\affiliation{Department of Physics and Astronomy, Iowa State University, Ames, Iowa 50011, USA}

\author{Anton A. Burkov}
\affiliation{Department of Physics and Astronomy, University of Waterloo, Waterloo, Ontario, Canada N2L 3G1}

\author{Ruslan Prozorov}
\affiliation{Ames Laboratory, Iowa State University, Ames, Iowa 50011, USA}
\affiliation{Department of Physics and Astronomy, Iowa State University, Ames, Iowa 50011, USA}

\author{Sergey L. Bud'ko}
\affiliation{Ames Laboratory, Iowa State University, Ames, Iowa 50011, USA}
\affiliation{Department of Physics and Astronomy, Iowa State University, Ames, Iowa 50011, USA}

\author{Adam Kaminski}
\affiliation{Ames Laboratory, Iowa State University, Ames, Iowa 50011, USA}
\affiliation{Department of Physics and Astronomy, Iowa State University, Ames, Iowa 50011, USA}

\author{Paul C. Canfield}
\email[]{canfield@ameslab.gov}
\affiliation{Ames Laboratory, Iowa State University, Ames, Iowa 50011, USA}
\affiliation{Department of Physics and Astronomy, Iowa State University, Ames, Iowa 50011, USA}

\date{\today}

\begin{abstract}
Magnetic Weyl semimetals are expected to have extraordinary physical properties such as a chiral anomaly and large anomalous Hall effects that may be useful for future, potential, spintronics applications.\,\cite{Yan2017,zhang2019} However, in most known host materials, multiple pairs of Weyl points prevent a clear manifestation of the intrinsic topological effects. Our recent density functional theory (DFT) calculations study suggest that EuCd$_{2}$As$_{2}$ can host Dirac fermions in an antiferromagnetically (AFM) ordered state or a single pair of Weyl fermions in a ferromagnetically (FM) ordered state.\,\cite{Wang2019} Unfortunately, previously synthesized crystals ordered antiferromagnetically with $T_{\textrm{N}}$\,$\simeq$\,9.5\,K.\,\cite{Schellenberg2011} Here, we report the successful synthesis of single crystals of EuCd$_{2}$As$_{2}$ that order ferromagnetically (FM) or antiferromagnetically (AFM) depending on the level of band filling, thus allowing for the use of magnetism to tune the topological properties within the same host. We explored their physical properties via magnetization, electrical transport, heat capacity, and angle resolved photoemission spectroscopy (ARPES) measurements and conclude that EuCd$_{2}$As$_{2}$ is an excellent, tunable, system for exploring the interplay of magnetic ordering and topology.
\end{abstract}

\maketitle 
To date, a number of magnetic topological materials have been proposed. GdPtBi\,\cite{canfield1991} is a proposed magnetic field driven Weyl semimetal.~\cite{hirschberger2016} Multiple Weyl points were found in canted antiferromagnetic state of YbMnBi$_{2}$.~\cite{Borisenko2019} Furthermore, interesting topological features have been observed in some ferromagnetic Kagome lattice materials including Co$_{3}$Sn$_{2}$S$_{2}$,~\cite{liu2018} Fe$_{3}$Sn$_{2}$~\cite{ye2018} and FeSn.~\cite{kang2019} To be more specific, theoretical predictions suggested three pairs of Weyl points in Co$_{3}$Sn$_{2}$S$_{2}$ with out-of-plane ferromagnetic order, and a giant anomalous Hall effect and ARPES results support this scenario.~\cite{Lin2012,Xu2018,liu2018,liu2019magnetic} Fe$_{3}$Sn$_{2}$ has two Dirac cones with 30\,meV gap near the Fermi level.~\cite{ye2018} A flat band and a pair of Dirac bands were observed in FeSn.~\cite{kang2019} However, a material with a single pair of Weyl points that readily offers tuning of its topological states is yet to be found. 

Recently, the layered triangular lattice compound EuCd$_{2}$As$_{2}$ was identified as a possible AFM Dirac semimetal with a single Dirac cone located close to the Fermi level when its in-plane, three fold symmetry is preserved.\,\cite{Hua2018} Subsequently, density functional theory (DFT) calculations on EuCd$_{2}$As$_{2}$ predicted that a FM ordered state with Eu moments aligned out-of-plane can split the Dirac cone into a single pair of Weyl points.\,\cite{Wang2019} However, previous experimental studies on EuCd$_{2}$As$_{2}$ show an A-type AFM below $T_{N}$\,$\simeq$\,9.5\,K that consists of FM triangular layers stacked antiferromagnetically, with moments pointing in the layer.\,\cite{Schellenberg2011, Wang2016,Rahn2018} Unfortunately, the in-plane three fold symmetry is broken in this spin configuration. In this case, the Dirac cone is no longer protected and a gap opens.\,\cite{Rahn2018} Very recently, an ARPES study above the AFM transition temperature claimed that the effective breaking of time reversal symmetry by FM-like fluctuations, associated with strong intralayer FM correlations of the A-type AFM order, can induce Weyl nodes.~\cite{ma2019} All of these results point toward the importance of stabilizing a FM state in EuCd$_{2}$As$_{2}$.   

Here, by discovering and taking advantage of the chemical tunability of EuCd$_{2}$As$_{2}$, we report the successful growths of single crystals of EuCd$_{2}$As$_{2}$ with two different magnetic ground states: EuCd$_{2}$As$_{2}$ that orders magnetically at low temperatures with a ferromagnetic component to its long range order (FM-EuCd$_{2}$As$_{2}$) and EuCd$_{2}$As$_{2}$ that orders magnetically at low temperatures without any detectable ferromagnetic component (AFM-EuCd$_{2}$As$_{2}$). Whereas for many local moment compounds, the desire to tune or change an AFM state to a FM state can be considered to be an unachievable pipe-dream, in some cases, chemical substitutions, or even just widths of formation, can be used to accomplish just this feat. For example in the simple, binary, CeGe$_{2-x}$ system, depending on the value of $x$, there can be either a FM or AFM transitions.\,\cite{budko2014} Based on our DFT calculations,\,\cite{Wang2019} we expect to have a topological insulator in the AFM-EuCd$_{2}$As$_{2}$ and two pairs of Weyl points in the FM-EuCd$_{2}$As$_{2}$. In addition, a single pair of Weyl points can be further realized with a magnetic field applied along the crystallographic $c$ direction. (see SI) 

Single crystals of both FM-EuCd$_{2}$As$_{2}$ and AFM-EuCd$_{2}$As$_{2}$ were grown via solution growth using a salt mixture as flux. The difference in growth procedure between FM-EuCd$_{2}$As$_{2}$ and AFM-EuCd$_{2}$As$_{2}$ was the initial stoichiometry of Eu:Cd:As in the salt mixture. We also grew single crystals of EuCd$_{2}$As$_{2}$ using Sn flux and these crystals also manifest AFM order. We confirmed the crystal structure and composition via X-ray diffraction patterns and scanning transmission electron microscopy (STEM) with energy dispersive spectroscopy (EDS). (more details of crystal growth and experiments can be found in SI) 

\begin{figure}[!ht]
	\includegraphics[width=6in]{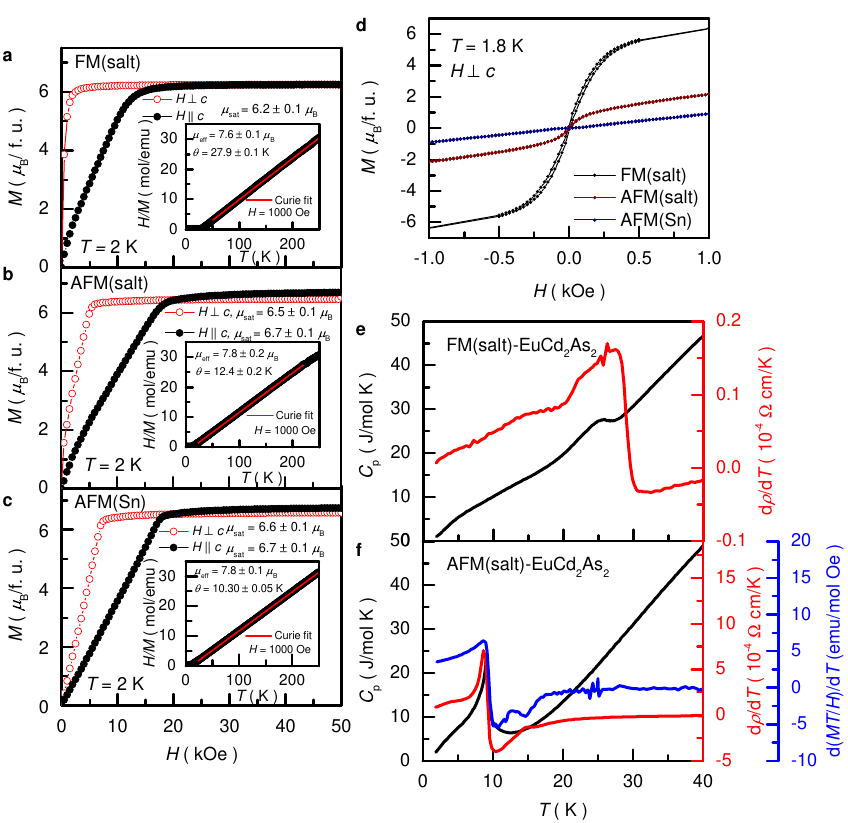}%
	\caption{\textbf{Specific heat, resistivity and magnetization data for FM-EuCd$_{2}$As$_{2}$ and AFM-EuCd$_{2}$As$_{2}$}  (a)-(c) Magnetic field dependent magnetization with the field parallel to the $c$ axis (black filled circle) and perpendicular to the $c$ axis (red open circle) at 2\,K for FM-EuCd$_{2}$As$_{2}$, AFM-EuCd$_{2}$As$_{2}$, and Sn flux grown AFM-EuCd$_{2}$As$_{2}$ respectively. Inset shows the inverse susceptibility with Curie-Weiss fitting. (d) A comparison between the magnetic field dependent magnetization for the three samples, FM-EuCd$_{2}$As$_{2}$, AFM- EuCd$_{2}$As$_{2}$, and Sn flux grown AFM-EuCd$_{2}$As$_{2}$, in the low field regime between -1000\,Oe to 1000\,Oe. All the measurements shown were done with the field perpendicular to the $c$-axis. (e) Temperature dependent specific heat $C_{P}$ (black line, left axis) and resistivity derivatives (red line, right axis) for FM-EuCd$_{2}$As$_{2}$. (f) Temperature dependent specific heat $C_{P}$ (black line, left axis), resistivity derivatives (red line, right axis) and d($MT/H$)/d$T$ at $H_{\parallel c}$\,=\,50\,Oe (blue line, the second right axis) for AFM-EuCd$_{2}$As$_{2}$.}
	\label{fig:MTRT}
\end{figure}

In order to determine the transition temperatures and nature of the magnetic ground state in these crystals, we conducted specific heat, resistivity, and magnetization measurements. Figures\,\ref{fig:MTRT} (a)-(c) show the anisotropic $M(H)$ and $H/M(T)$ data for the three representative crystals we have studied: FM(salt)-EuCd$_{2}$As$_{2}$, AFM(salt)-EuCd$_{2}$As$_{2}$, and AFM(Sn)-EuCd$_{2}$As$_{2}$. At $T$\,=\,2\,K, each of these samples becomes saturated by roughly 20\,kOe for $H\,\parallel\,c$; for $H\,\perp\,c$, the $M(H)$ data saturates at progressively lower and lower fields as we progress from AFM(Sn) to AFM(salt) to FM(salt). All samples displays a magnetic easy axis that lies within the layers. Whereas the $H\,\perp\,c$ data in Fig.\,\ref{fig:MTRT} (a) strongly suggests a FM state, Figure\,\ref{fig:MTRT} (d) shows that, indeed, the FM(salt) sample develops an unambiguous remnant field at $H$\,=\,0 for full, four-quadrant $M(H)$ loops. Other differences between these samples include FM(salt) having a resolvably lower $\mu_{sat}$ and $\mu_{eff}$ and higher Curie Weiss theta value than the AFM samples. These data suggest that FM(salt) samples have less than the full Eu$^{2+}$ occupancy of the Eu sites. This is consistent with multiple powder X-ray data sets we have collected and analyzed. Both lab-based as well as synchrotron based data indicate that the FM(salt) sample have Eu vacancies at the several percent level. (see S.I.) At a finer level of comparison, the AFM(salt) sample is closer to the AFM(Sn) sample, but intermediate in its $M(H)$ behavior at 1.8\,K and Curie Weiss temperature value. This is an observation that we will return to once we present our ARPES data in Fig.\,\ref{fig:arpes} below. 

We can also compare the behavior of FM(salt) and AFM(salt) near their respective transition temperatures. FM(salt) has a broader, and much higher temperature, feature than AFM(salt) in the specific heat data as shown in Fig.\,\ref*{fig:MTRT} (e) and (f), respectively. Using the peak as a criterion, the transition temperature for FM(salt) is $T_{\textrm{C}}\,\simeq\,26.4$\,K, and the transition temperature for AFM(salt) is $T_{\textrm{N}}\,\simeq\,9.2$\,K. In addition to these transitions, both EuCd$_{2}$As$_{2}$ samples have a broad shoulder at temperatures below $T_{N}$ of $T_{C}$. The origin of this additional anomaly can be attributed to the thermal population of the 4f crystal-field levels that are split by the molecular field acting on Eu ions.~\cite{Marquina1996}  

Figures\,\ref{fig:MTRT} (e) and (f) also show temperature dependent resistivity derivatives (d$\rho$/d$T$) of FM(salt) and AFM(salt) samples. ($\rho(T)$ data shown in SI.) Near the magnetic transition temperature, d$\rho$/d$T$ is found to resemble the specific heat.\,\cite{Fisher1968} Clear signatures of a phase transition are observed for both FM(salt) and AFM(salt). Figure\,\ref{fig:MTRT} (f) also shows the temperature dependent $\frac{\textrm{d}(MT/H)}{\textrm{d}T}$ at $H$\,=\,50\,Oe along the crystallographic $c$ axis on AFM(salt) which also reveals a feature similar to that seen in the specific heat data;\,\cite{Fisher1962} this analysis is formally only appropriate for AFM transitions (not FM ones) and is not shown in Fig.\ref{fig:MTRT} (e).

\begin{figure}
	\includegraphics[width=3in]{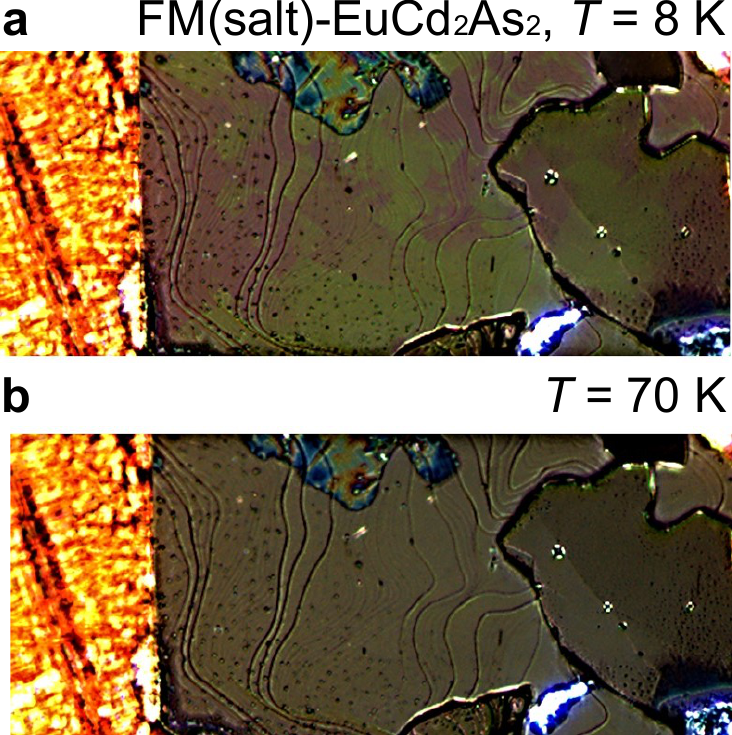}%
	\caption{\textbf{Magneto optical image of FM(salt)-EuCd$_{2}$As$_{2}$} (a) Ferromagnetic domains below $T_{\textrm{C}}$ imaged at $T$\,=\,8\,K. (b) Absence of domains above the transition, imaged at $T$\,=\,70\,K. Orange area on the left image is copper sample holder.}
	\label{fig:mo}
\end{figure}

To confirm the FM nature of FM(salt) samples, magneto-optical images were taken at temperatures above and below the transition temperature (see Fig.\,\ref{fig:mo} as well as SI), and comparison reveals the formation of magnetic domains below the transition. Similar imaging was performed on AFM(salt) but no domains were detected at any temperature. (see SI) In addition to observation of hysteresis, the ferromagnetic domains seen in Fig.\,\ref{fig:mo} (and studied in further detail in SI) provide a second, clear, indication that there is a net ferromagnetic component to the ordered state in FM(salt). On the other hand, the data shown in Fig.\,\ref{fig:MTRT} (d) and SI Fig.S7 are consistent with an antiferromagnetic ground state of AFM(salt).

\begin{figure}[!ht]
	\includegraphics[width=6.5in]{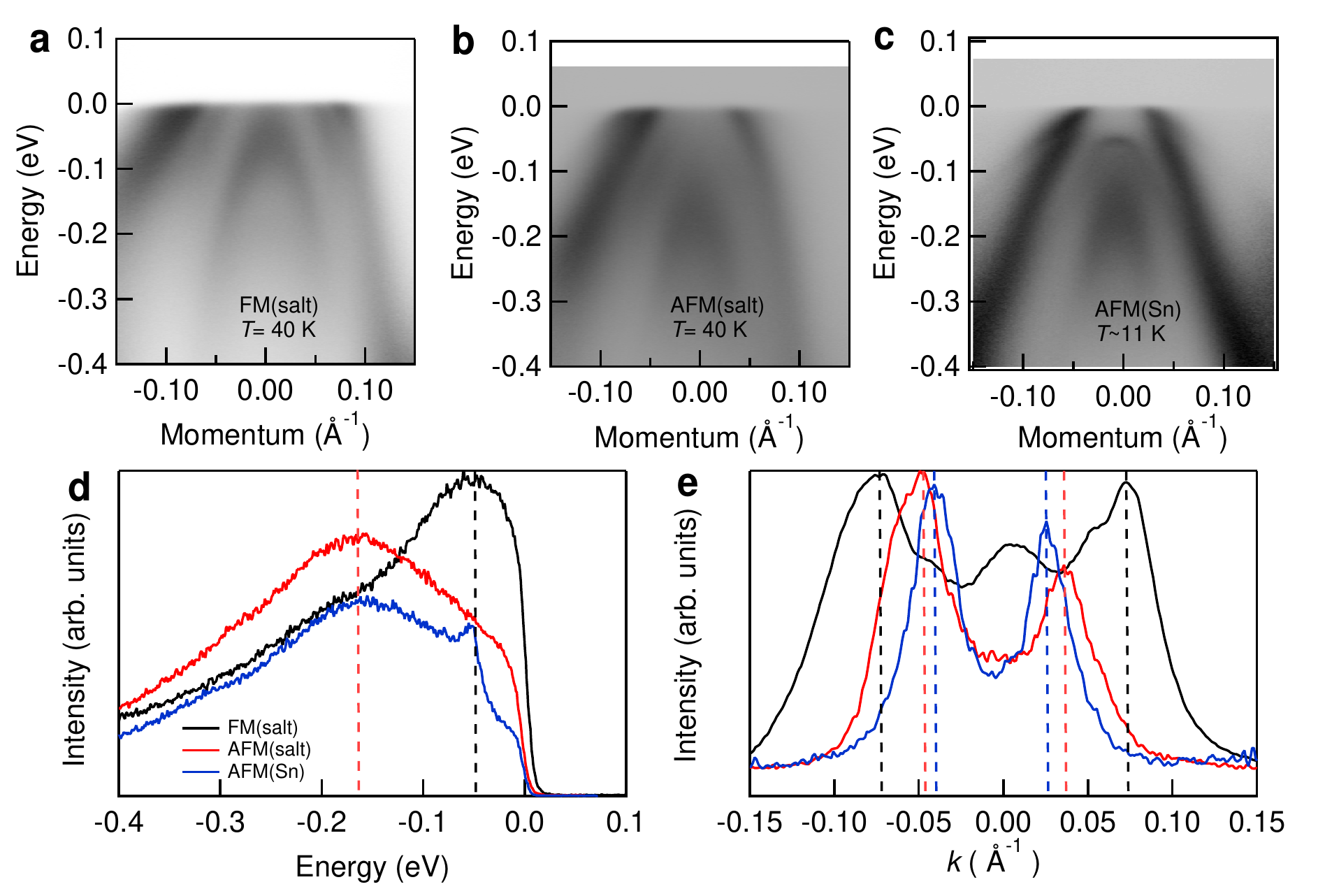}%
	\caption{\textbf{ARPES measurement of band structure of FM(salt)-EuCd$_{2}$As$_{2}$, AFM(salt)-EuCd$_{2}$As$_{2}$, and AFM(Sn)-EuCd$_{2}$As$_{2}$.} ARPES intensity plot along a cut through Gamma point for: (a) FM(salt) at $T$\,=\,40\,K. (b) AFM(salt) at $T$\,=\,40\,K. (c) Electronic structure of AFM(Sn) at $T\,\sim\,$11\,K. (d) Energy distribution curve at the zero momentum of FM(salt), AFM(salt), and AFM(Sn) with black, red and blue lines respectively. Dashed lines indicate the energy of top of the inner hole band for FM(salt) and AFM(salt). (e) Momentum distribution curves (MDC) at $E_{F}$ of FM(salt), AFM(salt), and AFM(Sn) with black, red and blue line respectively. Dashed lines (black for FM(salt), red for AFM(salt), and blue for AFM(Sn)) are indicating the peak positions of MDCs that mark the value of the Fermi momentum.}
	\label{fig:arpes}
\end{figure}

In order to probe the possible electronic origins that distinguish the FM(salt), AFM(salt) and AFM(Sn) samples, ARPES measurements were performed at low temperature. (See Fig.\,\ref{fig:arpes}) The data clearly show changes in the band filling: there is an almost rigid band shift of the hole pocket that crosses $E_{F}$ to higher binding energy as we progress from FM(salt) to AFM(salt) to AFM(Sn). Energy distribution curves at the $\Gamma$ point shown in Fig.\,\ref{fig:arpes} (d) also demonstrate that the top of the inner hole band of FM(salt) is $\sim$\,120\,meV higher than the one of AFMs. In addition, we observe a smaller size of the pocket in momentum distribution curve at the $E_{F}$. (See Fig.\,\ref{fig:arpes} (e)) The difference in the band filling between FM(salt), AFM(salt), and AFM(Sn) could be associated with either Eu-site occupancy or the ratio of divalent to trivalent (non-magnetic) Eu. 

\begin{figure}[!ht]
	\includegraphics[width=6.5in]{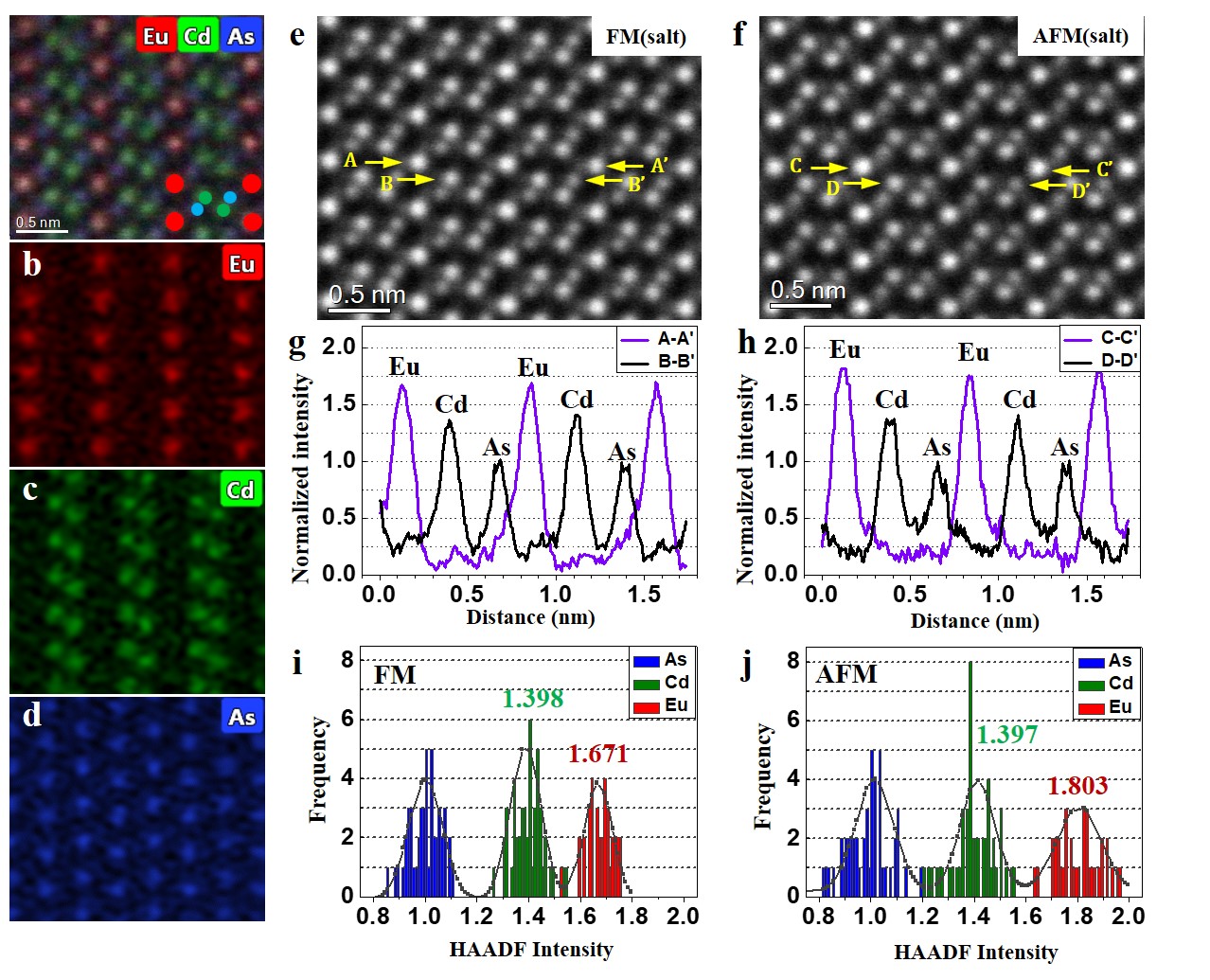}%
	\caption{\textbf{HAADF STEM and EDS analysis of FM(salt)-EuCd$_{2}$As$_{2}$ and AFM(salt)-EuCd$_{2}$As$_{2}$ along [21 $\bar{1}$ $\bar{0}$] zone axis.} (a) High resolution HAADF-STEM image of FM(salt) superimposed with color composite EDS elemental maps, and overlaid atomic model. (b)-(d), Atomically resolved EDS map of Eu, Cd and As taken from the same area of (a). (e)-(f) High resolution HAADF image of FM(salt) and AFM(salt), respectively. (g)-(h) Line profiles showing the image intensity (normalized to equal one for As column) as a function of position in image (e) along A-A' and B-B', and positions in image (f) along C-C' and D-D', respectively. (i)-(j), Histogram of the intensities of atomic image maxima in the area of (e) and (f), respectively.}
	\label{fig:TEM}
\end{figure}

To better correlate the differences in magnetization and band filling between FM(salt) and AFM(salt) with possible composition differences, we performed STEM on both FM(salt) and AFM(salt). The results suggest more Eu vacancies in FM(salt). Color composite, high-resolution high-angle-annular-dark-field (HAADF) scanning transmission electron microscopy (STEM) image and EDS elemental mapping of FM(salt) along the [21 $\bar{1}$ $\bar{0}$] crystallographic direction (Figs.\,\ref{fig:TEM} (a)-(d)) clearly demonstrate individual Eu, Cd and As atomic columns. To reveal any possible Eu and Cd site occupancy difference in AFM(salt) and FM(salt) samples, we directly compared the intensity of Eu, Cd and As columns in HAADF images of both samples (Figs.\,\ref{fig:TEM} (e) and (f)).\,\cite{krivanek2010} The images were taken under the same experimental conditions and sample thickness ($\sim$\,10\,nm). Figs.\,\ref{fig:TEM} (g) and (h) show line profiles through the two locations marked in Figs.\,\ref{fig:TEM} (e) and (h), respectively. The intensities were normalized to equal one for As column. Profile A-A' and C-C' indicate the intensity of Eu columns, and profile B-B' and D-D' show the intensity of Cd-As pairs. The Eu column of FM(salt) shows slightly lower intensity ($\sim$\,1.7) than that of AFM(salt) ($\sim$\,1.8), whereas Cd column indicates almost the same intensity. Figures\,\ref{fig:TEM} (i) and (j) show a histogram of the normalized peak intensities (As peak as 1) for all the atoms in the Figs.\,\ref{fig:TEM} (e) and (f), respectively. A theoretical Gaussian fit\,\cite{krivanek2010} for the distributions of the different species of the atoms, based on the standard deviations determined experimentally for the Eu, Cd and As atoms is overlaid on the figure. It is clearly demonstrated that the Eu columns of FM(salt) has lower average intensity (1.671) than that of AFM(salt) (1.803), suggesting more vacancies in FM(salt). The TEM results is also consistent with X-ray results shown in SI. In addition, the presence of Eu vacancies in FM samples, which will lower the electron count, is also consistent with ARPES data indicating a lower $E_{F}$. 

Our results show that EuCd$_{2}$As$_{2}$ is a rare material that can be tuned from having an antiferromagnetic ground state to a ferromagnetic one. By changing growth conditions, EuCd$_{2}$As$_{2}$ can be shifted from a $T_{N}$\,$\simeq$\,9.2\,K antiferromagnet to a $T_{C}$\,$\simeq$\,26.4\,K ferromagnet. This change in ground state is associated with a clear shift in the electronic structure as well as measured Eu$^{2+}$ content. Detailed DFT calculations predict that AFM state is a host to topological insulator, while FM state hosts two pairs of Weyl points. This can be further tuned to one pair of Weyl points by polarizing spins along the crystallographic $c$ direction. This material is therefore an ideal candidate for studies of the interplay of magnetism and topology and the macroscopic manifestation of Weyl Fermions.

\newpage

\begin{acknowledgements}
The authors thank A. Bhattacharya and R. McDonald for helpful discussion. This work was supported by the U.S. Department of Energy, Office of Basic Energy Sciences, Division of Materials Sciences and Engineering. The research (N.H.J., Y.W., E.T., A.P., D.H.R., K.L., B.S., R.P., S.L.B., P.C.C) was performed at Ames Laboratory. Ames Laboratory is operated for the U.S. Department of Energy by the Iowa State University under Contract No. DE-AC02-07CH11358. This work was also supported by the Center for Advancement of Topological Semimetals, (N.H.J., B.K., L.L.W, B.G.U., R.J.M., A.A.B., A.K.) an Energy Frontier Research Center funded by the U.S. Department of Energy Office of Science, Office of Basic Energy Sciences, through the Ames Laboratory under its Contract No. DE-AC02-07CH11358. Use of the Advanced Photon Source at Argonne National Laboratory was supported by the U.~S.~Department of Energy (DOE), Office of Science, Office of Basic Energy Sciences, under Contract No.~DE-AC$02$-$06$CH$11357$. T.K. and L.Z. are supported by Laboratory Directed Research and Development funds through Ames Laboratory. All TEM and related work were performed using instruments in the Sensitive Instrument Facility in Ames Lab.  
\end{acknowledgements}

\clearpage

\section{Supplementary information}

\subsection{Methods}

Single crystals of EuCd$_{2}$As$_{2}$ were grown using salt (NaCl/KCl) as flux. We followed a procedure qualitatively similar to that describe in Ref.\,\onlinecite{Schellenberg2011} with quantitative modifications: Eu (Ames Laboratory, 99.99+\,$\%$), Cd (Alfa Aesar, 99.9997~$\%$) and As (Alfa Aesar, 99.9999~$\%$) were weighed in the molar ratio of 1:2:2 for FM(salt)-EuCd$_{2}$As$_{2}$ and 1.75:2:2 for AFM(salt)-EuCd$_{2}$As$_{2}$, and mixed with fourfold mass ratio of an equimass mixture of NaCl (Alfa Aesar, ultra dry, 99.99~$\%$) and KCl (Alfa Aesar, ultra dry, 99.95~$\%$). All the elements and salts were weighted and put into a fused silica tube in a glove box with an Ar atmosphere. We minimized the air exposure by attaching a Swagelok ball-valve to the tube when we move it from the glove box to glass-sealing bench. After sealing the first fused silica ampoule, we put it into the second fused silica tube, which has slightly larger diameter, with silica wool supporting it on both ends. The nested ampoules were then heated to 469\,$^{\circ}$C over 25 hours, held for 24 hours, then heated to 597\,$^{\circ}$C and held for 24 hours, and finally heated up to 847\,$^{\circ}$C, and subsequently slowly cooled to 630\,$^{\circ}$C over 44 hours. We took the ampoule out of the furnace at 630\,$^{\circ}$C and quenched in air, in order to avoid single crystal growth of the binary, CdAs$_{2}$, phase. The resulting EuCd$_{2}$As$_{2}$ crystals were hexagonal plates with the crystallographic $c$ axis perpendicular to the crystal surface as shown in Fig.\,S1 (g). Sn flux grown single crystals of EuCd$_{2}$As$_{2}$ were grown with following procedure. An initial stoichiometry of Eu:Cd:As:Sn = 1:2:2:10 was put into a fritted alumina crucible (CCS)\,\cite{Canfield2016} and sealed in fused silica tube under a partial pressure of argon. The thus prepared ampoule was heated up to  $900 ^\circ$C over 24 hours, and held there for 20 hours. This was followed by a slow cooling to $500 ^\circ$C over 200 hours, and decanting of the excess flux using a centrifuge. 

Magnetization measurements were performed in a Quantum Design, Magnetic Property Measurement System, SQUID magnetometer. The samples were mounted on a Poly-Chloro-Tri-Fluoro-Ethylene disk, and the separately measured background was subtracted. The magnetization values for all three samples are calculated using molecular weight corresponding to the stoichiometric EuCd$_{2}$As$_{2}$. 

Temperature dependent electrical transport measurements were carried out in a Quantum Design Physical Property Measurement System for temperature range of 2\,K\,$\le\,T\,\le$\,300\,K. The samples for electrical transport measurements were prepared by attaching four Pt wires using Epotek-H20E silver epoxy and DuPont 4929N silver paint. The current was applied in $ab$ plane with $I\,=\,$1\,mA and $f\,=\,$17\,Hz. Temperature dependent specific heat measurements on conglomerates of EuCd$_{2}$As$_{2}$ crystallites, using a hybrid adiabatic relaxation technique of the heat capacity option in a Quantum Design, Physical Property Measurement System.

Temperature magneto-optical Kerr-effect imaging was performed using a helium flow-type cryostat with the ability to apply a magnetic field.\,\cite{Kreyssig2009,Tanatar2009} Linearly polarized light optical imaging was done using a Leica DMLM microscope equipped with high quality polarizer and analyzer. In the experiment, the sample is positioned on a gold-plated copper cold finger with its flat surface perpendicular to the light propagation direction. The light polarization direction, also perpendicular to the light direction, was controlled by a polarizer and could be changed with respect to the stationary sample.

In the magnetically ordered phase, the magnetic permittivity depends on the angle between the polarization direction of the incoming light and the direction of spontaneous magnetization which results in ferromagnetic domains of different orientations having different brightness and color contrast when observed through an analyzer oriented almost perpendicular to the polarizer axis. The small angle is required to induce asymmetry of the appearance of the domains of different orientation. No contrast is expected in the antiferromagnetic state where microscopic magnetization averages to zero at the scale of a wavelength, about 600 nm.

Samples used for ARPES measurements were cleaved in situ at 40\,K under ultrahigh
vacuum (UHV). The data were acquired using a tunable VUV laser ARPES system, that
consists of a Omicron Scienta DA30 electron analyzer, a picosecond Ti:Sapphire oscillator and fourth harmonic generator.\,\cite{Rui2014} Data were collected with photon energies of 6.7\,eV. Momentum and energy resolutions were set at $\sim$\,0.005\,$\textrm{\AA}^{-1}$ and 2\,meV. The diameter of the photon beam on the sample was $\sim$\,30\,$\mu$m.

TEM specimens were prepared by focused ion beam system (FIB, FEI Helios NanoLab G3). Both samples were lifted-out from crystal surface that cleaved right before sample preparation. The FM(salt) and AFM(salt) samples were prepared at the same FIB session and loaded to the same TEM grid to minimize the effect of surface oxidation. Scanning transmission electron microscopy (STEM) was performed on a probe aberration corrected Titan Themis STEM equipped with a Super-X EDS detector. STEM analysis was conducted at 200\,kV using a sub-angstrom electron probe with a convergence angle of 18 mrad and a detection angle of 74-200\,mrad. High-resolution high angle annular dark field (HAADF) images were acquired by using the drift corrected frame integration (DCFI) function in Velox software (Thermo Fisher) to improve the image quality with reducing shifts and instabilities. The individual frames were taken with a dwell time of 800 ns and a probe current of $\sim$\,20\,pA. The 40 frames were aligned and combined. Electron energy loss spectroscopy (EELS) was performed using a Gatan Quan ER 965 with a collection angle of 26 mrad.

\subsection{DFT calculations}

\begin{figure}[!ht]
	\includegraphics[width=6in]{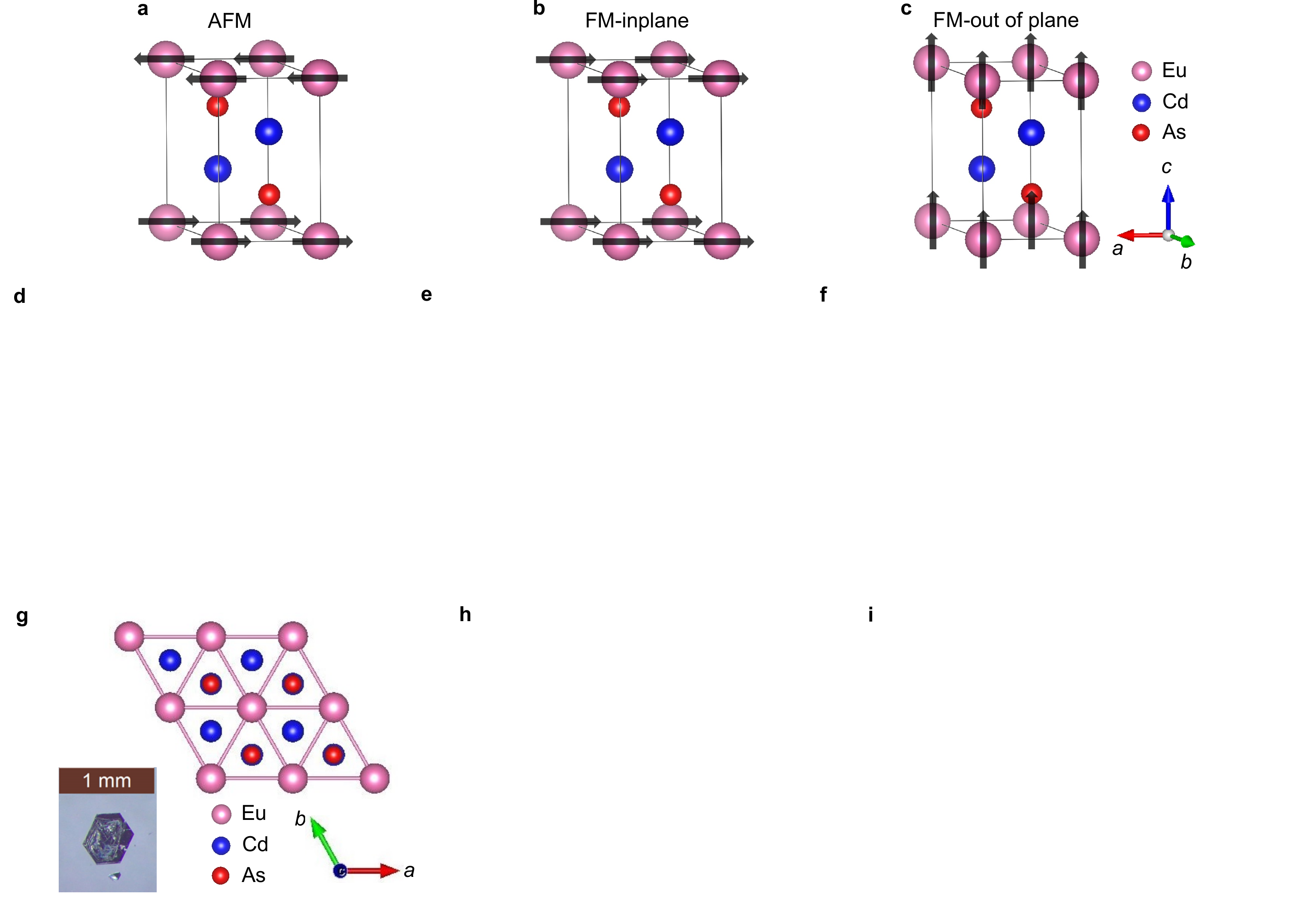}%
	\caption{\textbf{DFT calculation results of AFM-EuCd$_{2}$As$_{2}$, in-plane FM-EuCd$_{2}$As$_{2}$ and out-of-plane FM-EuCd$_{2}$As$_{2}$} Spin structure of \textbf{a}, AFM, \textbf{b}, in-plane FM, and \textbf{c}, out-of-plane FM. Bulk band structures along $k_{z}$ near $\Gamma$ point of different magnetic configurations: \textbf{d}, AFM, \textbf{e}, in-plane FM, and \textbf{f}, out-of-plane FM. The green shade shows the projection on As $p$ orbitals. \textbf{g}, Triangular symmetry of Eu sites in in-plane ($ab$ plane) and picture of the single crystal of EuCd$_{2}$As$_{2}$. \textbf{h}, Fermi arc of the in-plane FM configuration on (001) surface. Weyl points projection on the arc plots, +1 in red and -1 in blue \textbf{i}, Fermi arc of the out-of-plane FM configuration on (110) surface. Weyl points projection on the arc plots, +1 in red and -1 in blue}
	\label{fig:Spin}
\end{figure}

Band structure with spin-orbit coupling (SOC) in DFT\,\cite{Hohenberg1964,Kohn1965} has been calculated with the PBE7 exchange-correlation functional, a plane-wave basis set and projected augmented wave method\,\cite{Bloechl1994} as implemented in VASP.\,\cite{Kresse1996,Kresse1996a} To account for the half-filled strongly localized Eu 4$f$ orbitals, a Hubbard-like\,\cite{Dudarev1998} U value of 5.0\,eV is used. For bulk band structures of FM-EuCd$_{2}$As$_{2}$, a Monkhorst-Pack\,\cite{Monkhorst1976} (11 11 7) $k$-point mesh with a Gaussian smearing of 0.05\,eV including the $\Gamma$ point and a kinetic energy cutoff of 318\,eV have been used. For AFM-EuCd$_{2}$As$_{2}$, the hexagonal unit cell is doubled along $c$ direction with a (11 11 3) $k$-point mesh. Experimental lattice parameters\,\cite{Schellenberg2011} have been used with atoms fixed in their bulk positions. A tight-binding model based on maximally localized Wannier functions\,\cite{Marzari1997,Souza2001,Marzari2012} was constructed to closely reproduce the bulk band structure including SOC in the range of $E_{F}\,\pm\,1$\,eV with Eu $sdf$, Cd $sp$ and As $p$ orbitals. Then the 2D Fermi surface, including the Fermi arcs of semi-infinite surfaces, were calculated with the surface Green’s function methods\,\cite{Lee1981,LeeJoannopoulos1981,Sancho1984,Sancho1985} as implemented in Wannier Tools.\,\cite{Wu2018}

Based on our DFT calculations, shown in Fig.\,\ref{fig:Spin}, we expected to have a self-doped topological insulator in AFM-EuCd$_{2}$As$_{2}$ (Fig.\,\ref{fig:Spin} (a) and (d)) and two pairs of Weyl points in FM-EuCd$_{2}$As$_{2}$ (Fig.\,\ref{fig:Spin} (b), (e) and (h)). In addition, a single pair of Weyl points can be further realized with the applied magnetic field along the crystallographic $c$ direction. (Fig.\,\ref{fig:Spin} (c), (f) and (i))  

The band inversion around the $\Gamma$ point between the conduction band (derived from Cd $s$ orbital) and valence band (derived from As $p$ orbitals) is evident in Fig.\,\ref{fig:Spin} (d) for AFM-EuCd$_{2}$As$_{2}$. Each band is doubly degenerate due to the effective time reversal symmetry (ETRS)\,\cite{Mong2010} $S$\,=\,$\Theta \tau_{1/2}$, where $\Theta$ is the time reversal operation and $\tau_{1/2}$ is the non-symmorphic translation along $c$. When the AFM moment is in-plane, breaking the 3-fold rotation symmetry, the band inversion gives a self doped strong topological insulator state, AFM-TI or AFM-TCI.

For FM, the ETRS is broken. Thus, the double degeneracy of each band is lifted as seen in Fig.\,\ref{fig:Spin} (e) and (f). When the FM moment is in-plane and along $a$ direction (Fig.\,\ref{fig:Spin} (e)), or $k_{y}$ direction, the conduction and valence bands are almost touching along $k_{z}$ near the $\Gamma$ point. Each of the DP is split into two Weyl points (WPs) along $k_{y}$ direction at (0.000, $\pm$\,0.002, $\pm$\,0.017 1/$\AA$; 0.011\,eV). When projected on (001) surface, the WPs of the same chirality from these two pairs of WPs overlap and give Fermi arcs that merge into a circular Fermi surface as shown in Fig.\,\ref{fig:Spin} (h). Interestingly, when the FM moment is out-of-plane and along $c$ direction (Fig.\,\ref{fig:Spin} (f)), or $k_{z}$ direction, one pair of WPs split from the DPs are pushed over the $\Gamma$ point and annihilated, whereas the other pair survives and is pushed away from $\Gamma$ point, thus resulting in the case of a single pair WPs at (0.000, 0.000, $\pm$\,0.030 1/$\AA$; 0.048\,eV). The Fermi arc of such single pair of WPs when projected on (110) surface is shown in Fig.\,\ref{fig:Spin} (i). 

\subsection{X-ray diffraction pattern and elemental analysis}

\subsubsection{Benchtop x-ray experiments}

\begin{figure}[!ht]
	\includegraphics[scale=1]{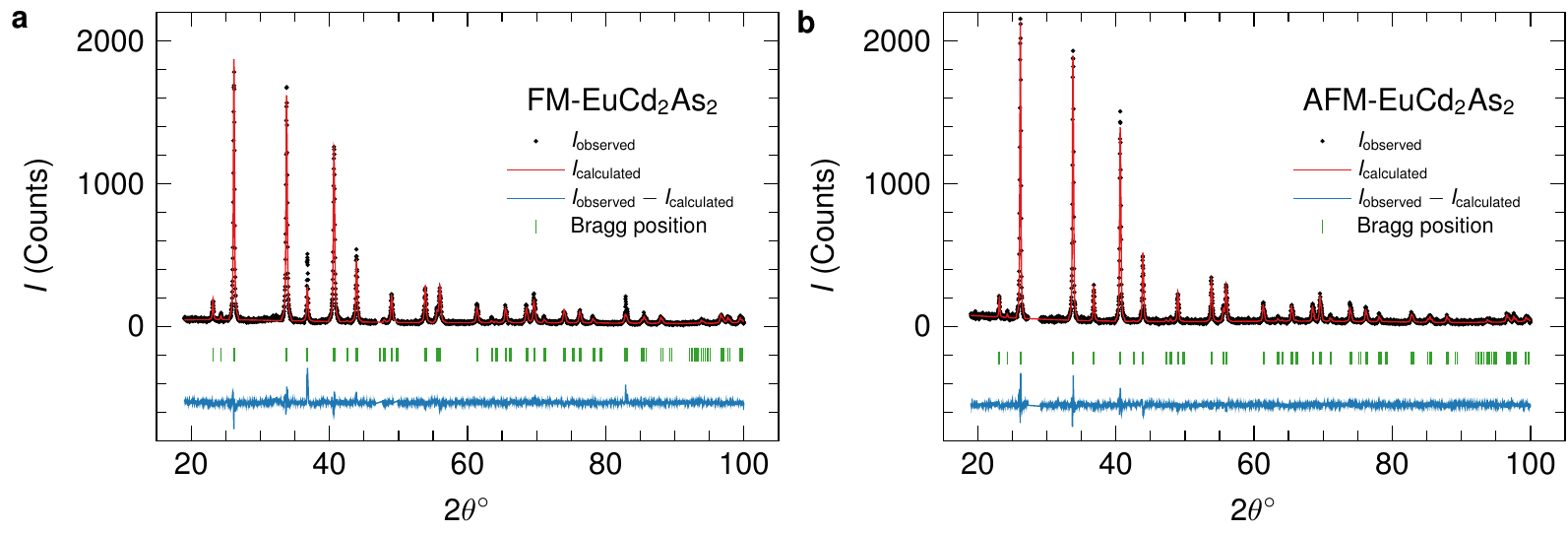}%
	\caption{Room temperature benchtop x-ray diffraction data for powder samples of ferromagnetic \textbf{a} and antiferromagnetic \textbf{b} EuCd$_{2}$As$_{2}$. Lines show the fits made via Rietveld refinements and the difference between the data and the fit. Tick marks indicate Bragg peak positions.
		\label{XRD}}
\end{figure}

A Rigaku MiniFlex diffractometer (Cu $K_{\alpha1}$ and $K_{\alpha2}$ radiation) was used for acquiring x-ray diffraction (XRD) patterns at room temperature. Data are shown in Fig.~\ref{XRD} along with fits made via Rietveld refinements using \textsc{fullprof}\,\cite{Rodriguez1993}. Table~\ref{Tab_lab_xray} lists the parameters determined from the refinements, which support the fact that the Eu crystallographic site has partial occupancy of $0.96(1)$ for the ferromagnetic sample. Note that for the antiferromagnetic sample, allowing the Eu site occupancy to vary generated non-physical (means over 100\,$\%$) results. Hence, the occupancy was fixed at $1$. Similarly, the occupancy of the Cd and As sites could not be varied for either sample. Results for the antiferomagnetic sample are similar to previously published results, although the values for the thermal parameter $B$ are somewhat larger than previously reported: $B_{\text{Eu}}=1.08(6)$, $B_{\text{Cd}}=1.27(7)$, and $B_{\text{As}}=1.09(7)$\,\cite{Rahn2018}. These results are qualitatively similar to our analysis of effective and saturated moments based on the data shown in Figs.\,1 (c), (d); i.e. FM(salt)-EuCd$_{2}$As$_{2}$ has a reduced Eu occupancy relative to AFM(salt)-EuCd$_{2}$As$_{2}$.

\begin{table}
	\caption{Parameters determined from Reitveld refinements of room temperature X-ray diffraction data taken on powder samples of FM(salt) and AFM(salt) EuCd$_{2}$As$_{2}$. Refinements were made using space group P$\bar{3}$m1 with Eu at the $1a~(0,0,0)$ Wyckoff position, and Cd and As at the $2d~(1/3,2/3,z)$ Wyckoff position. \label{Tab_lab_xray}}
	\begin{ruledtabular}
		\begin{tabular}{@{\hspace{0em}} c c c p{1.3cm} @{\hspace{0em}}  }
			\multicolumn{1}{@{\hspace{3em}} c}{}&\multicolumn{1}{ c}{FM(salt)-EuCd$_2$As$_2$}&\multicolumn{1}{@{\hspace{2em}}c }{AF(salt)-EuCd$_2$As$_2$}\\
			\colrule
			$a$~(\text{\AA}) & 4.4365(2)& 4.4398(2)\\
			$c$~(\text{\AA})  & 7.3247(4)& 7.3277(4)\\
			\text{Eu Site Occupancy} & 0.96(1)  & 1.00 \\
			$z_{\text{Cd}}$ & 0.6342(5) & 0.6339(5) \\
			$z_{\text{As}}$ & 0.2515(8) & 0.2459(8) \\
			B$_{\text{Eu}}$~(\text{\AA}$^2$) & 2.4(3) & 3.0(2) \\
			B$_{\text{Cd}}$~(\text{\AA}$^2$) & 2.1(2) & 2.8(2) \\
			B$_{\text{As}}$~(\text{\AA}$^2$) & 2.0(2) & 2.1(2) \\
			$\chi^{2}$ & 1.89 & 1.69 \
		\end{tabular}
	\end{ruledtabular}
\end{table}

\subsubsection{High-energy x-ray experiments}
High-resolution high-energy x-ray powder diffraction (HEXRD) data were collected at the 11-BM beamline \cite{Wang_2008} at the Advanced Photon Source, Argonne National Laboratory, using an average x-ray wavelength of $\lambda=0.45786$~\AA\ ($E=27.079$~keV). An analyzer system comprised of twelve perfect Si($111$) analyzers and twelve LaCl$_3$ scintillators spaced $2\degree$ apart in scattering angle $2\theta$ \cite{Lee_2008} were employed along with discrete detectors covering a $22$\degree\ angular range.  During the measurements, the detectors were scanned at a speed of $0.01$\degree$/$s over a $34$\degree\ range with data points collected every $0.001$\degree.  This process resulted in data covering $2\theta\approx0.5$ to $50$\degree.  A mixture of NIST standard reference materials, Si (SRM $640$c) and Al$_2$O$_3$ (SRM $676$) were used to calibrate the instrument, where the Si lattice constant was used to determine the wavelength for each detector. Corrections were applied for detector sensitivity, $2\theta$ offset, small differences in wavelength between detectors, and the source intensity, as noted by an ion chamber, before merging the data into a single set of intensities evenly spaced in $2\theta$.  During measurements, the sample was spun at $\approx90$~Hz after mounting by a Mitsubishi robotic arm \cite{Toby_2009} and cooled down to $T=100$~K using an Oxford Cryosystems Cryosteam Plus device.

\begin{figure*}[]
	\centering
	\includegraphics[width=1.0\linewidth]{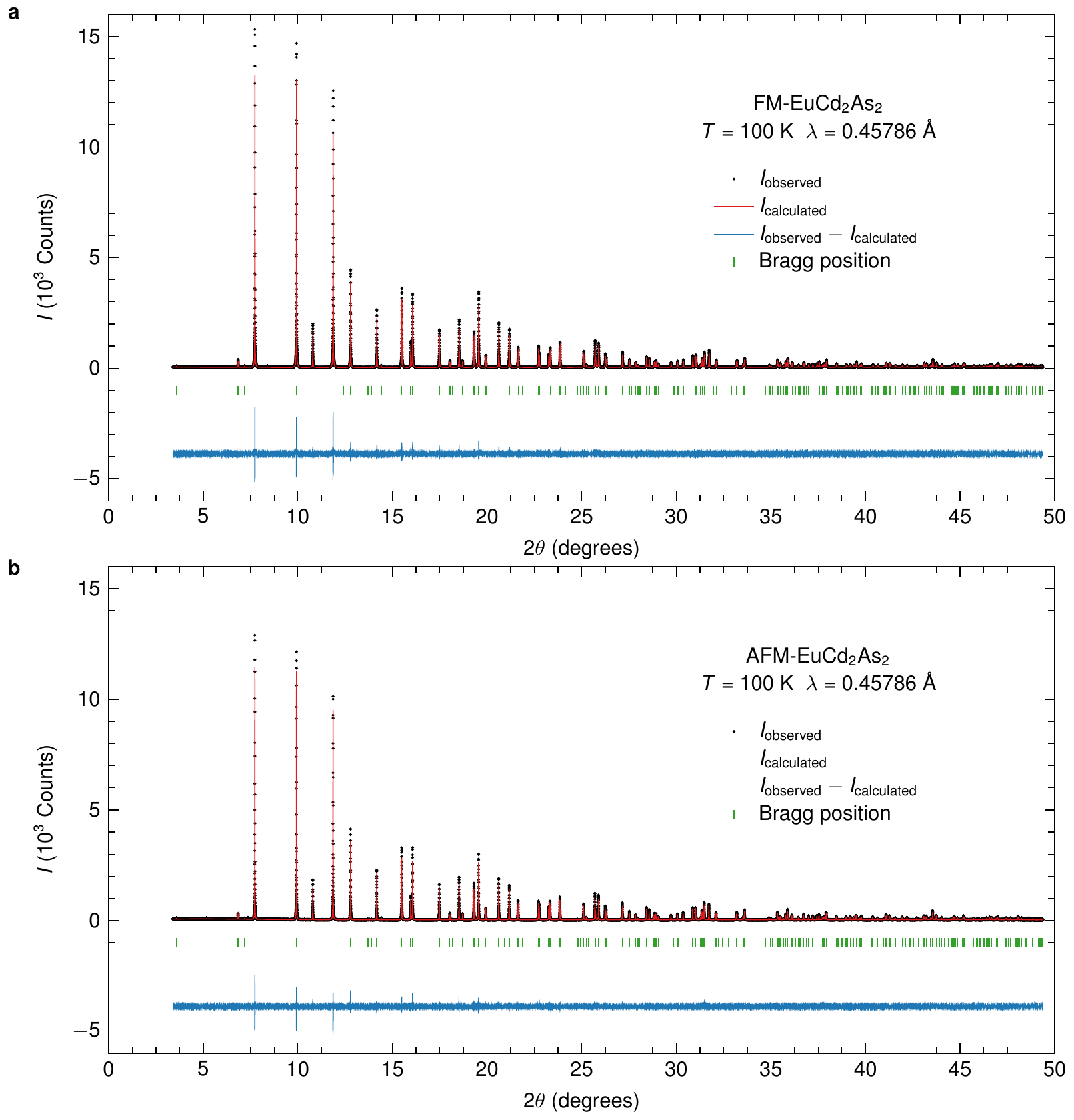}
	\caption{High-energy x-ray diffraction data for powder samples of ferromagnetic \textbf{a} and antiferromagnetic \textbf{b} EuCd$_{2}$As$_{2}$ cooled down to $T=100$~K. For better fitting, the peaks associated with impurities or apparent artifacts were removed. Lines show the fit made via Reitveld refinements and the difference between the data and the fit. Tick marks indicate Bragg peak positions.\label{Fig_aps_xray} }
\end{figure*}

Results from Reitveld refinements made using \textsc{gsas} \cite{Larson_2004} with \textsc{expgui} \cite{Toby_2001} are presented in Fig.~\ref{Fig_aps_xray} and Table~\ref{Tab_aps_xray}. A correction for x-ray absorption and appropriate anomalous dispersion coefficients were included in the refinements.   In contrast to the benchtop XRD data, the occupancy for all three sites could be refined, and the Eu crystallographic site of the AF sample is found to have full occupancy.  On the other hand, similar to the benchtop x-ray results, the HEXRD show that the Eu site of the FM sample is only partially occupied.   Thus, the room temperature benchtop XRD and $T=100$~K HEXRD data indicate that for a given temperature both the AF and FM samples show similar crystallographic parameters with the significant exception being less than full occupancy of the Eu site for the FM sample. It should be noted that comparison of Eu occupancy based on X-ray or STEM analysis with Eu$^{2+}$ occupancy based on magnetization data suggest that there may well be some degree of Eu$^{3+}$ present in all samples.

\begin{table}
	\caption{Parameters determined from Reitveld refinements of high-energy ($E=27.079$~keV) x-ray diffraction data taken on powder samples of FM(salt) and AFM(salt)-EuCd$_{2}$As$_{2}$ at $T=100$~K.  Refinements were made using space group $P\bar{3}m1$ with Eu at the $1a~(0,0,0)$ Wyckoff position, and Cd and As at the $2d~(1/3,2/3,z)$ Wyckoff position. \label{Tab_aps_xray}}
	\begin{ruledtabular}
		\begin{tabular}{@{\hspace{0em}} c c c p{1.3cm} @{\hspace{0em}}  }
			\multicolumn{1}{@{\hspace{3em}} c}{}&\multicolumn{1}{ c}{FM(salt)-EuCd$_2$As$_2$}&\multicolumn{1}{@{\hspace{2em}}c }{AF(salt)-EuCd$_2$As$_2$}\\
			\colrule
			\rule{0pt}{3ex}$a~(\text{\AA})$ & 4.43021(1)& 4.43219(1)\\
			$c~(\text{\AA})$  & 7.30240(2)& 7.30436(1)\\
			$\text{Eu Site Occupancy}$ & 0.989(4)  & 1.001(4) \\
			$\text{Cd Site Occupancy}$ & 0.999(4)  & 0.998(4) \\
			$\text{As Site Occupancy}$ & 1.001(6)  & 1.002(4) \\
			$z_{\text{Cd}}$ & 0.63332(6) & 0.63289(7) \\
			$z_{\text{As}}$ & 0.24660(8) & 0.24741(9) \\
			$B_{\text{Eu}}~(\text{\AA}^2)$ & 0.07(1) & 0.02(1) \\
			$B_{\text{Cd}}~(\text{\AA}^2)$ & 0.12(1) & 0.08(1) \\
			$B_{\text{As}}~(\text{\AA}^2)$ & 0.07(1) & 0.05(1) \\
			$\chi^{2}$ & 1.961 & 1.615 \
		\end{tabular}
	\end{ruledtabular}
\end{table}

\begin{figure}[!ht]
	\includegraphics[width=4 in]{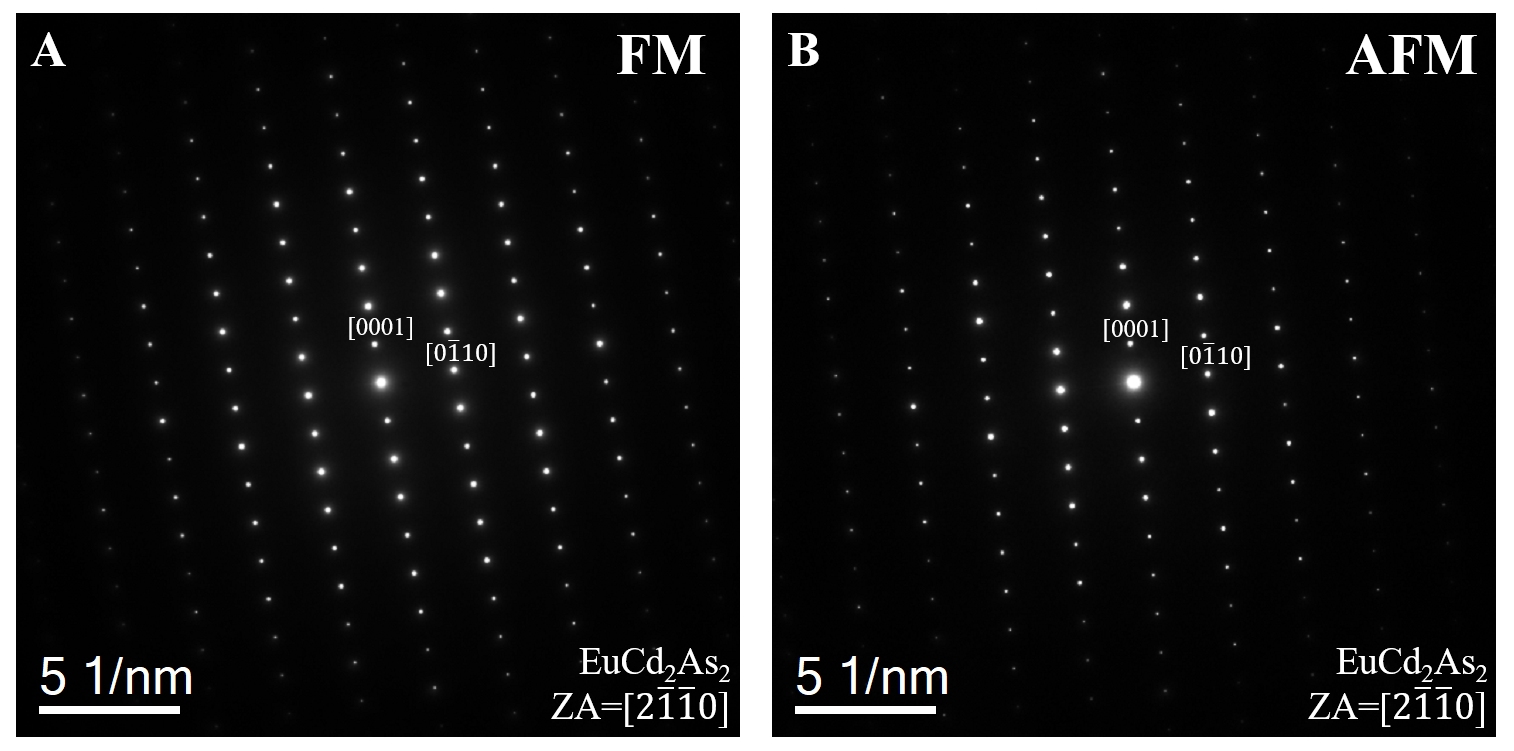}%
	\caption{\textbf{Electron diffraction patterns of FM(salt)-EuCd$_{2}$As$_{2}$ and AFM(salt)-EuCd$_{2}$As$_{2}$} Selected area Electron diffraction patterns taken from FM(salt)-EuCd$_{2}$As$_{2}$ and AFM(salt)-EuCd$_{2}$As$_{2}$ on [21 $\bar{1}$ $\bar{0}$] zone axis.}
	\label{fig:TEM2}
\end{figure}

In addition, FM(salt) and AFM(salt) samples shows the same hexagonal structure (hP5, P$\bar{3}$m1 space group), as confirmed by selected area electron diffraction patterns along the [21 $\bar{1}$ $\bar{0}$] zone axis (Fig.\,\ref{fig:TEM2}). 

\subsection{Temperature dependent resistivity}

\begin{figure}[!ht]
	\includegraphics[width=5in]{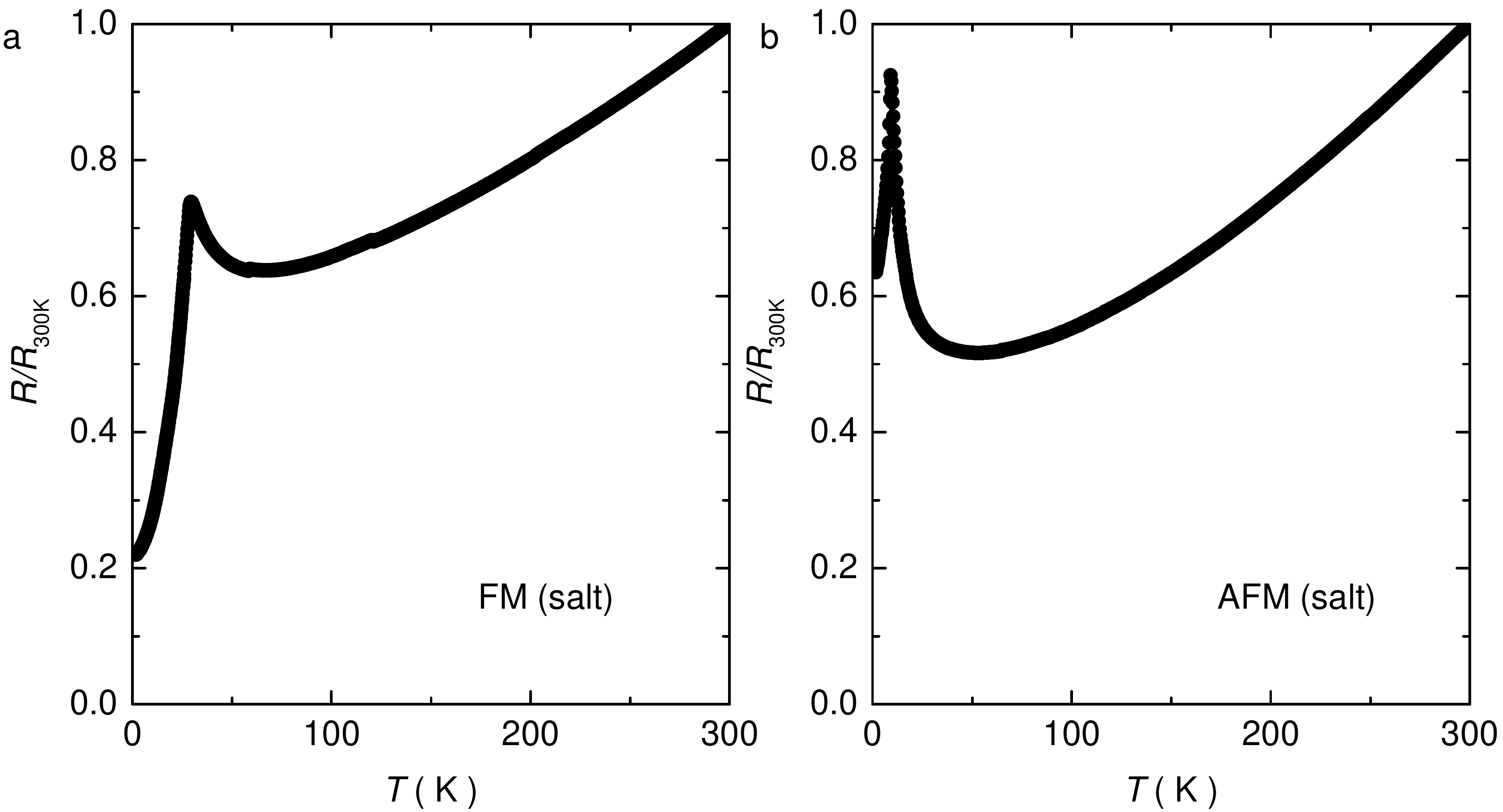}%
	\caption{\textbf{Temperature dependent resistivity} \textbf{a}, FM(salt)-EuCd$_{2}$As$_{2}$ and \textbf{b}, AFM(salt)-EuCd$_{2}$As$_{2}$.}
	\label{fig:SIRT}
\end{figure}

Figure\,\ref{fig:SIRT} shows temperature dependent resistivity of FM(salt)-EuCd$_{2}$As$_{2}$ and AFM(salt)-EuCd$_{2}$As$_{2}$ in zero applied magnetic field over the 2-300\,K range. These are the data used to determine $d\rho/dT$ plots shown in Figure.\,1. The resistivity upturn above the transition temperature (see Fig.\,\ref{fig:SIRT}) may be due to enhanced scattering on magnetic fluctuations.\,\cite{ma2019}

\subsection{Magneto optics}

\begin{figure}
	\includegraphics[width=3in]{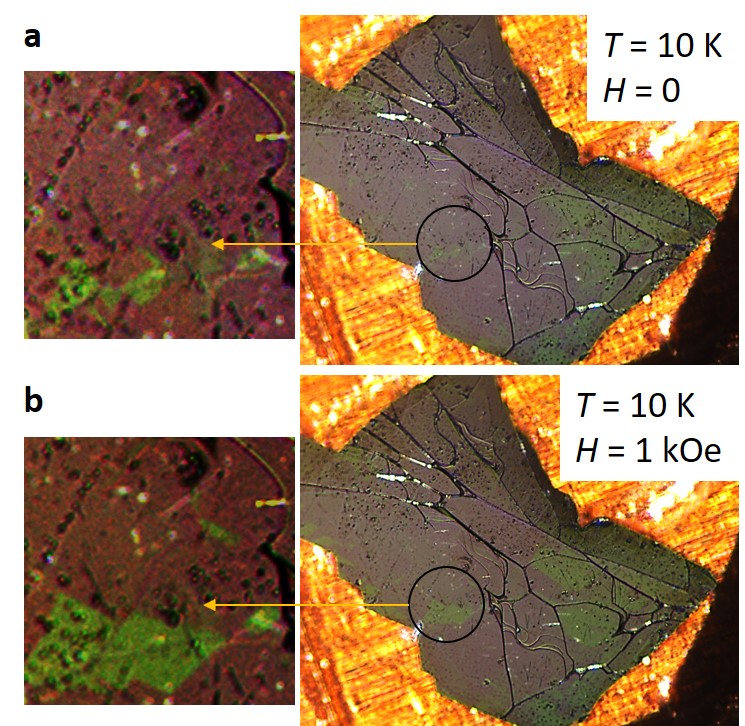}%
	\caption{\textbf{Applied magnetic field effect on domains of FM(salt)-EuCd$_{2}$As$_{2}$} Shift of magnetic domains by a transverse magnetic field (out of page) at $T$\,=\,10\,K. Circled area shows contrast-enhanced region shown on the left where the effect is particularly noticeable.  Orange areas around the sample shows copper substrate.}
	\label{fig:mo3}
\end{figure}

\begin{figure}
	\includegraphics[width=5in]{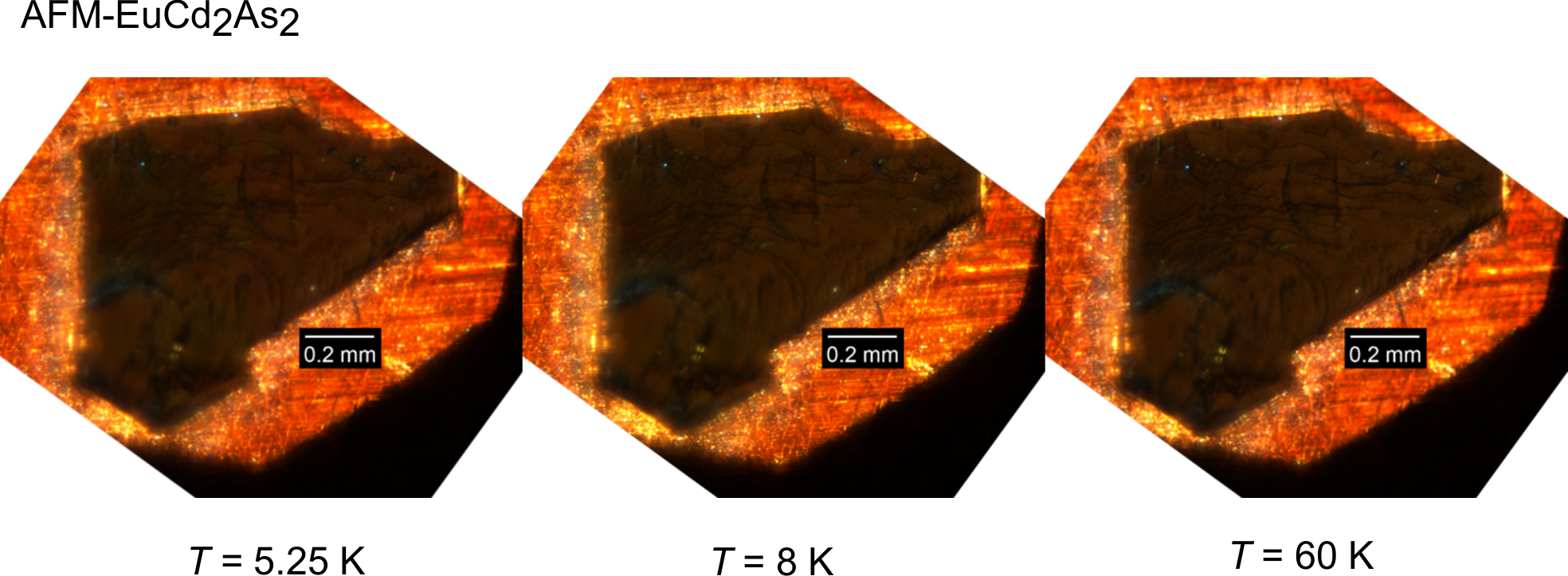}%
	\caption{\textbf{Magneto optical image of AFM(salt)-EuCd$_{2}$As$_{2}$} Absence of magnetic domains below $T_{\textrm{N}}$\,=\,9\,K. The image remains unchanged and shows no pattern above, just below and low temperatures with respect to $T_{\textrm{N}}$.}
	\label{fig:mo4}
\end{figure}

The ferromagnetic domains shown in Fig.\,2 can be manipulated by applied fields. Figure.\,\ref{fig:mo3} presents images showing the influence of a relatively weak magnetic field which is oriented perpendicular to the surface of the sample. In the circled area, we can see a domain growing in the 1\,kOe applied, transverse, field.

In contrast to the magneto optical images of FM(salt)-EuCd$_{2}$As$_{2}$ shown in Fig.\,2 and Fig.\,\ref{fig:mo3}, magneto optical images were taken of AFM(salt)-EuCd$_{2}$As$_{2}$ at temperatures above and below the transition temperature (see Fig.\,\ref{fig:mo4}), show that no domains were detected at any temperature. 

\clearpage

\section*{References}
\bibliographystyle{apsrev4-1}

%

\clearpage

\end{document}